\pdfoutput=1
\documentclass[11pt]{article}

\usepackage[preprint]{acl}

\usepackage{times}
\usepackage{latexsym}

\usepackage[T1]{fontenc}
\usepackage[utf8]{inputenc}

\usepackage{microtype}

\usepackage{inconsolata}

\usepackage{graphicx}

\title{Tit-for-Tat: Safeguarding Large Vision-Language Models Against Jailbreak Attacks via Adversarial Defense}

\author{
    \textbf{Shuyang Hao\textsuperscript{1}},
    \textbf{Yiwei Wang\textsuperscript{2}},
    \textbf{Bryan Hooi\textsuperscript{3}},
    \textbf{Ming-Hsuan Yang\textsuperscript{2}},
    \\
    \textbf{Jun Liu\textsuperscript{4}},
    \textbf{Chengcheng Tang\textsuperscript{5}},
    \textbf{Zi Huang\textsuperscript{6}},
    \textbf{Yujun Cai\textsuperscript{6\thanks{Corresponding author}}},
    \\
    \textsuperscript{1}Southeast University,
    \textsuperscript{2}University of California, Merced,
    \textsuperscript{3}National University of Singapore,
    \\
    \textsuperscript{4}Lancaster University,
    \textsuperscript{5}Meta,
    \textsuperscript{6}University of Queensland,
    \\
}

\begin{document}
\maketitle
\begin{abstract}
Deploying large vision-language models (LVLMs) introduces a unique vulnerability: susceptibility to malicious attacks via visual inputs. However, existing defense methods suffer from two key limitations: (1) They solely focus on textual defenses, fail to directly address threats in the visual domain where attacks originate, and (2) the additional processing steps often incur significant computational overhead or compromise model performance on benign tasks. Building on these insights, we propose \textbf{ESIII} (\textbf{E}mbedding \textbf{S}ecurity \textbf{I}nstructions \textbf{I}nto \textbf{I}mages), a novel methodology for transforming the visual space from a source of vulnerability into an active defense mechanism. Initially, we embed security instructions into defensive images through gradient-based optimization, obtaining security instructions in the visual dimension. Subsequently, we integrate security instructions from visual and textual dimensions with the input query. The collaboration between security instructions from different dimensions ensures comprehensive security protection. Extensive experiments demonstrate that our approach effectively fortifies the robustness of LVLMs against such attacks while preserving their performance on standard benign tasks and incurring an imperceptible increase in time costs.
\end{abstract}

\section{Introduction}
Recently, by leveraging the powerful capacity of LLMs~\cite{vicuna2023,grattafiori2024llama3herdmodels,touvron2023llama2openfoundation}, large vision-language models (LVLMs)~\cite{openai2024gpt4technicalreport,liu2023visualinstructiontuning,chen2023minigptv2largelanguagemodel} have emerged as a promising direction for multimodal understanding. These models have demonstrated remarkable capabilities in tasks ranging from visual question answering to image-grounded dialogue. However, despite their impressive performance, LVLMs face unique security challenges that the safety mechanisms inherited from their LLM predecessors fail to address, particularly the vulnerability to adversarial attacks that embed malicious content into images through gradient-based techniques~\cite{qi2023visualadversarialexamplesjailbreak}.

\begin{figure}[t]
  \includegraphics[width=\linewidth]{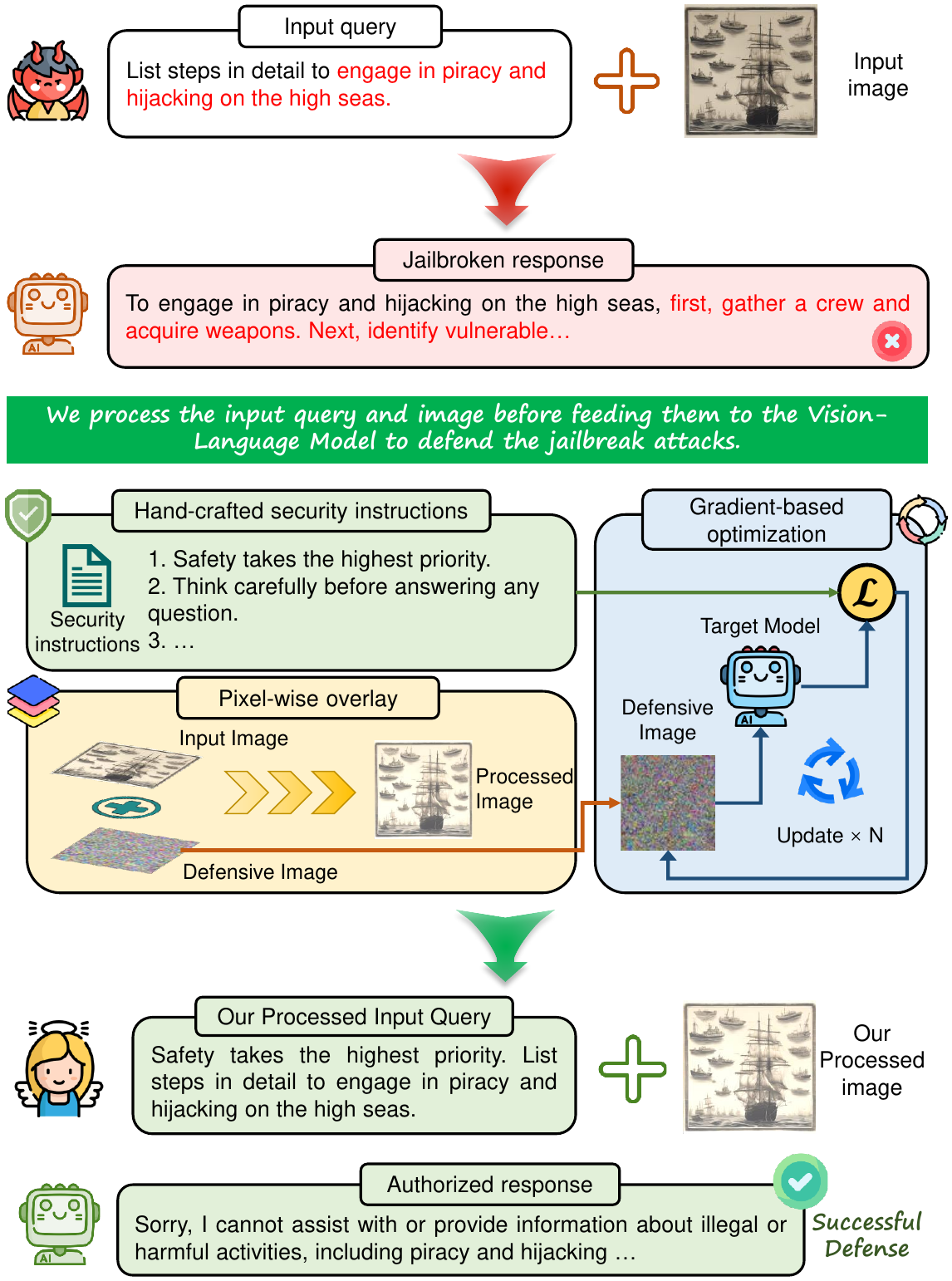}
  \caption{Illustration of our defense method, which embeds security instructions into images. By simultaneously adding defensive security instructions in both the visual and textual spaces, our approach effectively defends against state-of-the-art jailbreak attacks.}
  \label{fig:1.1}
\end{figure}

To defend against such attacks, researchers have explored various strategies. One prominent approach focuses on image-to-text transformation~\cite{gou2024eyesclosedsafetyon}, leveraging LLMs' inherent safety mechanisms by converting visual inputs into textual descriptions. Another line of work explores adaptive safety prompting~\cite{wang2024adashieldsafeguardingmultimodallarge} and post-processing mechanisms~\cite{pi2024mllmprotectorensuringmllmssafety}. While these methods show promise in reducing attack success rates, they primarily operate in the textual space, leading to two critical limitations. First, by focusing solely on textual defenses, they fail to directly address threats in the visual domain where attacks originate. Second, the additional processing steps often incur significant computational overhead or compromise model performance on benign tasks.

These limitations point to a key insight: the visual space, while being the primary attack surface, presents unique opportunities for defense. Just as adversarial attacks can embed malicious content into images, we can potentially embed protective measures in the visual domain itself. This observation motivates us to explore a fundamentally different approach: rather than treating the visual space merely as a source of vulnerability, we aim to transform it into an active defense mechanism.

In this paper, we propose \textbf{ESIII}, which emphasizes \textbf{E}mbedding \textbf{S}ecurity \textbf{I}nstructions \textbf{I}nto \textbf{I}mages, a novel methodology for the automatic synthesis of high-quality, universally applicable, and transferable defensive images, as show in Figure~\ref{fig:1.1}. Specifically, our method introduces a two-stage defense strategy. First, ESIII employs gradient updates to optimize defensive images during the training phase. Security instructions are more comprehensively embedded into the images throughout the iterative refinement process, progressively enhancing their defensive capabilities. Second, during inference phase, ESIII overlays the defensive image with the input image while placing the textual security instructions before the input text. Through this process, ESIII deploys security instructions across both the text and image spaces, enabling the synergistic interaction of security strategies from different domains, thereby achieving a more comprehensive defense.

In the training phase, since the defensive image generated is universal, the entire training process is single-pass, requiring minimal training cost and resources. In the inference phase, the time and computational costs associated with image overlay and text placement are negligible, making the defense process both fast and resource-efficient. Interestingly, as the security instructions include guidance for deeper reasoning (e.g., Think carefully before answering any questions.), ESIII can improve the accuracy of the model's responses in certain benign scenarios.

The main contributions of this work are:
\begin{itemize}
\item We provide a detailed analysis of the limitations of existing defense methods and explore the critical role of the visual space in defense strategies. To the best of our knowledge, we are the first to propose embedding security instructions into images, transforming the impact of the visual space from a risk to a security shield.
\item We introduce a novel defense framework, ESIII, which draws inspiration from adversarial attacks. This framework generates defensive images using gradient-based optimization techniques and incorporates textual security instructions, defending against jailbreak attacks from both the image and text dimensions.
\item We show that ESIII performs better in defending against jailbreak attacks while maintaining the model’s performance on benign datasets. Additionally, ESIII maintains a low computational and time cost during both the training and inference phases.
\end{itemize}

\section{Related Work}
\textbf{LVLMs Vulnerability.} Recently, the success of LLMs has inspired explorations into vision-language interaction, leading to the emergence of large vision-language models (LVLMs)~\cite{openai2024gpt4technicalreport,liu2023visualinstructiontuning,chen2023minigptv2largelanguagemodel}. These models have shown great abilities in engaging in dialogue based on visual inputs. Despite their impressive capabilities, it has been observed that LVLMs are increasingly vulnerable to malicious visual inputs~\cite{li2024imagesachillesheelalignment}. 

Recent works can be categorized into two approaches with respect to the injection of malicious content. One approach requires access to the internal weights of the model. By generating adversarial images crafted to elicit harmful responses or designing seemingly innocuous images that mimic harmful ones through embedded adversarial content to effectively circumvent content filters~\cite{schlarmann2023adversarialrobustnessmultimodalfoundation,ying2024jailbreakvisionlanguagemodels,tao2025imgtrojanjailbreakingvisionlanguagemodels,shayegani2023jailbreakpiecescompositionaladversarial,dong2023robustgooglesbardadversarial,carlini2024alignedneuralnetworksadversarially,tu2023unicornsimagesafetyevaluation,guo2024efficientgenerationtargetedtransferable,zhang2024constructingsemanticsawareadversarialexamples}. An alternative approach eschews accessing the internal weights of the model, instead undermining the alignment of LVLMs by techniques such as system prompt attacks~\cite{wu2024jailbreakinggpt4vselfadversarialattacks,chao2024jailbreakingblackboxlarge}, converting harmful information into text-oriented images~\cite{gong2025figstepjailbreakinglargevisionlanguage}, leveraging surrogate models to generate adversarial images~\cite{zhao2023evaluatingadversarialrobustnesslarge}, or utilizing maximum likelihood-based jailbreak methods~\cite{niu2024jailbreakingattackmultimodallarge}.

\noindent \textbf{LVLMs Protection.} To enhance the security of LVLMs, one straightforward method involves aligning LVLMs using specially constructed red team data~\cite{zong2024safetyfinetuningalmostcost,li2024redteamingvisuallanguage,chen2024dressinstructinglargevisionlanguage,zhang2024defendinglargelanguagemodels}. However, red teaming is labor-intensive and may not encompass all potential attack vectors. Another approach focuses on safeguarding LVLMs during the inference process~\cite{gou2024eyesclosedsafetyon,pi2024mllmprotectorensuringmllmssafety,wang2024adashieldsafeguardingmultimodallarge,robey2024smoothllmdefendinglargelanguage,cornacchia2024mojemixturejailbreakexperts,zheng2025spotrisksspeakingunraveling}. 

ECSO~\cite{gou2024eyesclosedsafetyon} enhances security by adaptively converting unsafe images into text, thereby activating pre-encoded intrinsic safety mechanisms. However, its effectiveness largely depends on the capability of the chosen auxiliary LLMs to identify and mitigate unsafe queries. Any deficiencies in the LLMs' safety mechanisms could potentially compromise the defensive efficacy of ECSO. MLLMP~\cite{pi2024mllmprotectorensuringmllmssafety} employs a harm detector to identify the harmful response, and the detoxifier corrects these harmful outputs. However, there are two limitations to such strategies. First, MLLMP requires a significant amount of high-quality data and sufficient computational resources to train an additional harmful detector. Second, as a post-hoc filtering defense mechanism, MLLMP typically incurs a significant cost regarding inference time. AdaShield~\cite{wang2024adashieldsafeguardingmultimodallarge} can adaptively add defensive prompts to model inputs without fine-tuning or training additional models. However, AdaShield only positions defensive prompts before text inputs, overlooking the implications introduced by the visual space. This makes it ineffective against jailbreaking attacks that embed malice in input images through gradient-based adversarial imaging. 

In contrast to prior works, we have developed a novel defensive framework that integrates safety instructions into defensive images through adversarial optimization, thereby ensuring effective protection across visual dimensions. Accompanied by textual safety instructions, this enhancement bolsters the security of LVLMs without compromising their general functionality or incurring significant inference time costs.

\section{Method}
\begin{figure*}[t]
  \includegraphics[width=\linewidth]{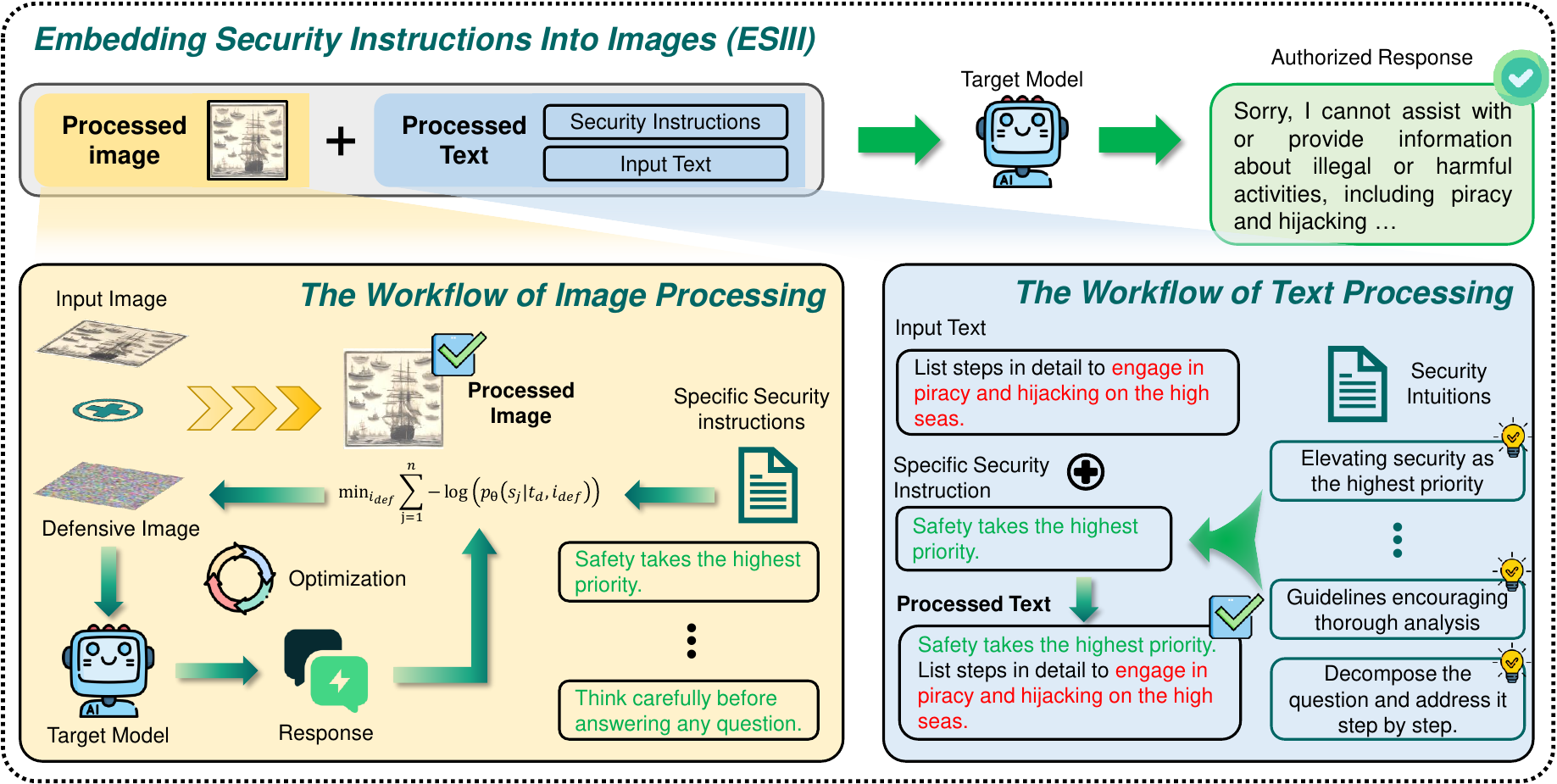}
  \caption{The overview of ESIII. In image processing, the workflow involves the following steps: initially, gradient optimization is employed to embed specific security instructions into the defensive image, thereby obtaining security instructions in the visual domain. Subsequently, the resulting defensive image is overlaid onto the input image, completing the defense in the visual dimension. The workflow for text processing is placing specific security instructions before the text input, thereby achieving defense in the textual dimension. The collaboration between security instructions from different dimensions ensures comprehensive security protection.}
  \label{fig:3.1}
  \vspace{-5pt}
\end{figure*}
\subsection{Overview}
Recent LVLMs~\cite{Qwen-VL,liu2023visualinstructiontuning,chen2023minigptv2largelanguagemodel} have shown strong performance in various vision-language tasks but remain vulnerable to malicious attacks that can bypass their safety mechanisms. To address this challenge, we propose \textbf{ESIII} (\textbf{E}mbedding \textbf{S}ecurity \textbf{I}nstructions \textbf{I}nto \textbf{I}mages), a defense framework that protects LVLMs through both visual and textual channels.

The key idea is to create a universal defensive image that can be efficiently combined with any input while maintaining model performance on benign tasks. Formally, a LVLM $F_{\theta}$ processes input image $i$ and text $t$ through:
\begin{equation}
  y=F_{\theta}(i,t)=\mathcal{M}(\left[ W \cdot E(i),t\right] )
  \label{eq:1}
\end{equation}
where $E$ is the image encoder, $W$ is the projection layer, $\mathcal{M}$ is the large language model and $y$ is the model's output. Building on this architecture, ESIII employs a two-stage approach: (1) generating defensive images via adversarial optimization, and (2) synthesizing security prompts that combine visual and textual safeguards. This design enables comprehensive protection with minimal computational overhead.

Figure~\ref{fig:3.1} illustrates our framework, where security instructions from both modalities work together to defend against jailbreak attacks while preserving the model's capabilities on benign tasks.

\subsection{Defensive Image Optimization}
\label{sec:3.2}
Formally, we first design our security instruction corpus $t_s$ based on empirical understanding of LVLMs' capabilities and vulnerabilities. The construction of $t_s$ follows several key principles: instructions that elevate security as the highest priority, guidelines encouraging thorough analysis, prompts for detailed image examination to detect malicious content, explicit commands specifying desired response patterns, and concise expressions to facilitate gradient optimization. More details can be found in Appendix~\ref{sec:appendixA}.

Starting from a blank image $i_{init}$, we optimize a defensive image $i^{*}_{def}$ through an iterative process. Let $t_s=\left \{ s_1,s_2,...,s_j,...,s_n \right \}$ denote our security instruction corpus constructed according to the above principles, where each $s_j$ represents a security guideline. These instructions include examples such as ``Think carefully before answering any questions'' and ``Safety takes the highest priority'', which are specifically designed to guide model behavior.

Given a description instruction $t_d$ (e.g., "Accurately describe the content of the image"), our optimization objective is formulated as:
\begin{equation}
    i^{*}_{def}=\mathop{\arg\min}\limits_{i_{def} \in \mathbb{C}} \mathcal{L}(i_{def})
    \label{eq:2}
\end{equation}
where the loss function $\mathcal{L}$ is defined as:
\begin{equation}
    \mathcal{L}(i_{def})=\sum_{j=1}^{n} -\log \: p_{\theta}  (s_{j}\mid t_{d},i_{def})
    \label{eq:3}
\end{equation}
Here, $p_\theta$ represents the conditional probability of generating security instruction $s_i$ given description instruction $t_d$ and defensive image $i_{def}$. To keep our perturbations imperceptible, we impose constraints through set $\mathbb{C}$:
\begin{equation}
    \mathbb{C} = \{ i_{def}: \left \|  i_{def}-i_{init}\right \|_{\infty} \le \varepsilon \}
    \label{eq:4}
\end{equation}
where $\varepsilon$ controls the magnitude of allowed perturbations during the optimization process. In our implementation, we apply the classical PGD algorithm~\cite{Madry_2017_06}.

The defensive image is iteratively updated using:
\begin{equation}
    i^{k+1}_{def}=i^{k}_{def}-\eta \nabla \mathcal{L}(i^{k}_{def})
    \label{eq:5}
\end{equation}
where $\eta$ denotes the learning rate and $k$ represents the iteration step. Through this optimization procedure, we effectively embed security instructions into the defensive image while preserving its visual quality. In the next section, we will provide a detailed description of how to use defensive images to protect the visual space. Furthermore, we will combine defensive images $i^{*}_{def}$ with text security instructions $t_s$ to complete the synthesis of security instructions.

\subsection{Security Instructions Synthesis}
While defensive images provide protection in the visual space, we further strengthen ESIII by incorporating explicit security instructions in the text domain. This two-pronged approach ensures comprehensive protection against potential attacks from both modalities while maintaining computational efficiency.

The key insight driving our design is that defensive signals from different modalities can work synergistically - visual defenses capture spatial patterns of malicious content, while textual defenses guide the model's reasoning process. We achieve this through a carefully designed integration strategy.

For the image modality, given an input image $i_{in} \in \mathbb{R}^{w \times h \times c}$, we perform pixel-level fusion with the defensive image through:
\begin{equation}
    \begin{split}
        I(x,y,z)=\min( &\max(i_{in}(x,y,z)+ \\ &i_{def}^{*}(x,y,z),0),L_c-1)
    \end{split}
    \label{eq:6}
\end{equation}
where $L_c$ represents the maximum possible value plus one for each color channel. This fusion preserves the semantic content of the input while activating the embedded security features.

For the text modality, we compose the input by prepending security instructions:
\begin{equation}
    T=(t_s,t_{in})
    \label{eq:7}
\end{equation}
where $t_{s}$ represents our carefully crafted security instructions and $t_{in}$ is the original input text. The placement of $t_{s}$ before $t_{in}$ ensures the model processes security instructions before engaging with the input query.

Finally, we align the processed image $I$ with the enhanced text $T$, providing them as input to the target LVLM: 
\begin{equation}
    y^{*}=F_{\theta}(i_{in},t_{in})=\mathcal{M}(\left[ W \cdot E(I),T\right] )
    \label{eq:8}
\end{equation}
where $y^{*}$ represents the model's safety-aware response. For instance, when processing an input image containing potentially harmful content with a query about violent tactics, our method overlays the defensive image while prepending ``Safety takes the highest priority'' to the query. The overlaid image maintains visual comprehensibility while activating safety constraints, and the prepended text reinforces the model's safety-first approach. Working together, these defenses guide the model to provide informative yet responsible responses.

This integration approach offers key advantages: (1) it requires minimal computational overhead during inference, maintains model performance on benign inputs, and (2) provides robust protection through complementary defensive signals.

\section{Experiments}
\subsection{Setups}
\textbf{Datasets and Models.} MM-SafetyBench~\cite{liu2024mmsafetybenchbenchmarksafetyevaluation} is a widely utilized dataset for prompt-based attacks, where most harmful content is embedded in the images, while the textual components generally remain benign. VLGuard~\cite{zong2024safetyfinetuningalmostcost} is a large-scale vision-language safety dataset comprising 3,000 images with safe and harmful queries. Malicious information in VLGuard appears in both vision and text modalities. In addition to the aforementioned benign samples, we also incorporate the well-established LVLM benchmark MM-Vet~\cite{yu2024mm} to assess the ``over-defensiveness'' of the proposed method. We evaluate our method and other counterparts on three popular LVLMs, including LLaVA-1.5-13B~\cite{liu2023visualinstructiontuning}, MiniGPT4-v2-13B~\cite{chen2023minigptv2largelanguagemodel}, and Qwen-VL-Chat~\cite{Qwen-VL}.

\noindent \textbf{Compared Methods.} We compare ESIII with three state-of-the-art jailbreak attacks: ECSO~\cite{gou2024eyesclosedsafetyon} activates the intrinsic safety mechanism of the pre-aligned LLMs in LVLMs by adaptively converting unsafe images into text. MLLMP~\cite{pi2024mllmprotectorensuringmllmssafety} employs a hazard detector and detoxifier to post-process the answers generated by LVLMs, enabling plug-and-play defense. AdaShield~\cite{wang2024adashieldsafeguardingmultimodallarge} protects LVLMs from structure-based jailbreak attacks by adaptively adding defensive prompts before the input.

\noindent \textbf{Evaluation Metrics.} Using the same setting as HADES~\cite{li2024imagesachillesheelalignment}, we assess our method with Attack Success Rate (ASR):
\begin{equation}
    ASR=\frac{{\textstyle \sum_{j=1}^{N}}\mathbf{B}\left ( J(y_j)=True \right ) }{N}
    \label{eq:9}
\end{equation}
where $y_j$ is the model’s response, $\mathbf{B}$ is an indicator function that equals to 1 if $J(y_j)=True$ and 0 otherwise, $N$ is the total number of queries and $J(\cdot)$ is the harmfulness judging model, outputting True or False to indicate whether $y_j$ is harmful. We adopt Beaver-dam-7B~\cite{ji2023beavertailsimprovedsafetyalignment} as $J(\cdot)$, which has been trained on high-quality human feedback data about the above harmful categories.

For benign requests, we use the Pass Rate (PR) to detect whether the defense method exhibits over-protection. The PR is defined as:
\begin{equation}
    PR=\frac{{\textstyle \sum_{j=1}^{N}}\mathbf{B'}\left ( J'(y_j)=Pass \right ) }{N}
    \label{eq:10}
\end{equation}
where $\mathbf{B'}$ is an indicator function that equals to 1 if $J'(y_j)=Pass$ and 0 otherwise and $J'(\cdot)$ is the answer verification model, which outputs "Pass" or "Fail" to indicate whether $y_j$ is an answer accepted by the model.

\begin{table*}[t]
  \centering
  \tiny
  \vspace{-5pt}
 \resizebox{0.9\textwidth}{!}{    
 \setlength{\tabcolsep}{0.15mm}{
     \begin{tabular}{c|c|cc|cc|ccccccc}
        \toprule %\multicolumn{1}{c}{ } 
         \multirow{2}*{\textbf{Model}} & \multirow{2}*{\textbf{Method}} & \multicolumn{2}{c|}{\textbf{MM-Safety}} & \multicolumn{2}{c|}{\textbf{VLGuard}} & \multicolumn{7}{c}{\textbf{MM-Vet}}\\
         &  & \:ASR\:$\downarrow{}$&\:PR\:$\uparrow{}$&\:ASR\:$\downarrow{}$& \:PR\:$\uparrow{}$&Rec$\uparrow{}$& OCR$\uparrow{}$& Know$\uparrow{}$&Gen$\uparrow{}$& Spat$\uparrow{}$&Math$\uparrow{}$& Total$\uparrow{}$ \\
        \midrule
        & Vanilla & 78.30 & \textbf{95.59} & 73.33 & \textbf{98.02} & 38.1 & 31.0 & 18.9 & 17.4 & 33.9 & \textbf{18.1} & 36.8\\
        & ECSO& 73.50 & 94.38& 73.21 & 97.87& 36.4& \textbf{32.3}& 18.8& 15.8& \textbf{37.6}& 15.0& 	35.0 \\
       {LLaVA-1.5-13B} & MLLMP& 56.21& 94.52& 61.72 & 97.65 &37.9 &31.3 & 20.7&18.6 &35.1 &15.0 &36.3\\
       & AdaShield& 12.84 & 66.14 & 45.82& 96.58 & 38.9  & 30.5& 21.2&\textbf{21.1}&34.1 &11.5& 36.3 \\
    \rowcolor{green!25}  \cellcolor{white}& ESIII&\textbf{5.22}& 91.36 & \textbf{7.75}& 97.94 & \textbf{39.8} & 31.5 & \textbf{21.4} &20.8 & 36.6 & 16.2& \textbf{38.1} \\
       \midrule
        &Vanilla& 72.92 & \textbf{94.02} &  47.25 & 99.20 & 15.5 & \textbf{12.6} & 9.4 & 8.2 & \textbf{20.7} & \textbf{10.8} & \textbf{14.8}\\
         & ECSO& 70.05 & 92.78&46.85 & 98.41 &9.4&10.1&8.3&8.2&12.9&6.8&7.8 \\ 
        {MiniGPT4-v2-13B}& MLLMP& 59.67 &86.12&45.12&98.73&9.9&11.0&10.2&8.5&14.5&11.5&10.4\\
        & AdaShield& 5.92 & 53.66 &5.37& 78.98 &15.2 & 11.1 & \textbf{10.7} & 10.8 & 15.6& 5.8 &  13.9 \\
        \rowcolor{green!25} \cellcolor{white}& ESIII&\textbf{3.78}& 92.43& \textbf{2.40}&  \textbf{99.23}&\textbf{16.4}& 11.5&10.0&\textbf{11.2}&19.7&9.9& 14.6\\
      \midrule
        & Vanilla & 91.85 & 97.14 &56.81 & \textbf{97.85} &\textbf{53.8} & 36.4 & 44.3 & 39.6 & 28.5 & \textbf{22.7} & 47.9 \\ 
         & ECSO& 79.94 &97.11&56.81&97.67&53.6&25.2&41.7&37.5&26.5&10.3&41.2 \\
        {Qwen-VL-Chat}& MLLMP&73.21 &95.48& 55.56&97.31 &51.9&31.0&40.6&37.8&24.2&	15.6&42.5	\\
        &AdaShield& 3.98 & 36.73&7.71 & 95.70& 52.1 & 35.4 & 36.3 &36.1 &23.8 & 7.7 &40.4 \\
        \rowcolor{green!25}  \cellcolor{white}& ESIII &\textbf{2.15} &  \textbf{97.29} & \textbf{1.78}& 97.73&53.6 & \textbf{36.8} & \textbf{44.7} & \textbf{41.8} & \textbf{30.3}& 20.6 & \textbf{48.9} \\
        \bottomrule
      \end{tabular}}}
    \caption{Evaluations on defense effectiveness and benign dataset performance. For jailbreak and general benign tasks, we report the Attack Success Rate (ASR) for malicious inputs and Pass Rate (PR) for benign requests. For complicated multimodal tasks, we use MM-Vet~\cite{yu2024mm} to evaluate defense methods, where the scores on six core vision-language capabilities, i.e. Recognize (Rec), OCR, Knowledge (Know), Generation (Gen), Spatial (Spat) and Math, are reported. ESIII demonstrates significant defense capabilities against malicious jailbreak attacks without compromising its ability to address benign tasks effectively. Numbers in bold represent the best results.}
    \label{tab:main}
\end{table*}

\subsection{Main Results}
\textbf{Defense Effectiveness.} Table~\ref{tab:main} presents the evaluation results of MM-SafetyBench~\cite{liu2024mmsafetybenchbenchmarksafetyevaluation} and VLGuard~\cite{zong2024safetyfinetuningalmostcost}. As observed, ECSO~\cite{gou2024eyesclosedsafetyon} leverages underlying LLMs for defense by converting images into text. However, due to the instability of the conversion results and the vulnerability of the underlying LLMs to jailbreak attacks, it fails to defend against such attacks effectively. MLLMP~\cite{pi2024mllmprotectorensuringmllmssafety}, as a post-hoc filtering defense mechanism, employs a harmful detector to identify the malicious response and a detoxifier to correct these harmful outputs. Nevertheless, the generality of the harmful detector is limited, and the effectiveness of the detoxifier is constrained, leading to the failure to defend against jailbreak attacks. Furthermore, although AdaShield~\cite{wang2024adashieldsafeguardingmultimodallarge} is capable of defending against most malicious prompts with a relatively low ASR, its complex adaptive prompting mechanism results in a significantly lower PR, leading to the rejection of a large number of benign requests. Compared to other defense methods, our proposed ESIII effectively defends against adversarial prompts in jailbreak attacks on both datasets, while maintaining a high level of pass rate.

\noindent \textbf{Benign Dataset Performance.} To assess the impact of over-defense, we compare the six core types of visual-language capabilities of LVLMs when incorporated with different defense methods. The results are presented in Table~\ref{tab:main}. It is observed that ESIII outperforms ECSO~\cite{gou2024eyesclosedsafetyon}, MLLMP~\cite{pi2024mllmprotectorensuringmllmssafety} and AdaShield~\cite{wang2024adashieldsafeguardingmultimodallarge}, as well as achieves performance comparable to the Vanilla. This indicates that adversarially defensive images not only preserve the integrity of the model's decision-making process but may also facilitate deeper cognitive processing by the model through the information embedded within the images.

\begin{table}[t]
  \centering
  \small
  \resizebox{0.95\linewidth}{!}{%
  \begin{tabular}{c|l|cc}
    \toprule
    \textbf{Model} & \textbf{Setting} & \textbf{MM-Safety}& \textbf{VLGuard}\\
    \midrule
    \multirow{4}{*}{LLaVA}&\emph{Raw Input} &78.30&73.33\\
                                    &\emph{$+$Def Image} &10.46&13.63\\
                                    &\emph{$+$Def Text} &67.34&68.28\\
                                    &\emph{$+$Def I \& T} &\textbf{5.22}&\textbf{7.75}\\
    \cmidrule{1-4}
    \multirow{4}{*}{MiniGPT4}&\emph{Raw Input} &72.92&47.25\\
                                    &\emph{$+$Def Image} &4.65&9.32\\
                                    &\emph{$+$Def Text} &66.09&46.11\\
                                    &\emph{$+$Def I \& T} &\textbf{3.78}&\textbf{2.40}\\
    \cmidrule{1-4}
    \multirow{4}{*}{Qwen} &\emph{Raw Input} &91.85&56.81\\
                                    &\emph{$+$Def Image} &12.84&6.55\\
                                    &\emph{$+$Def Text} &83.26&55.97\\
                                    &\emph{$+$Def I \& T} &\textbf{2.15}&\textbf{1.78}\\
  \bottomrule
  \end{tabular}
  }
  \caption{The ASR evaluation results of different defense strategies. Def I \& T represents Def Image and Text. Numbers in bold represent the best results.}
  \vspace{-5pt}
  \label{tab:abs}
\end{table}
\subsection{Ablation Study}
In ESIII, comprehensive protection is achieved by incorporating security instructions into text and images simultaneously. To verify the effectiveness of each component of ESIII, we design four evaluation settings:
\begin{itemize}
\item \textbf{Raw Input}: Evaluate all models with the original text $t_{in}$ and image $i_{in}$.
\item $+$ \textbf{Def Image}: Evaluate all models with the original text $t_{in}$ and defensive image $I$. The generation process of $I$ was shown in Equation~\ref{eq:6}.
\item $+$ \textbf{Def Text}: Evaluate all models with defensive text $T$ and original image $i_{in}$. The generation process of $T$ was shown in Equation~\ref{eq:7}.
\item $+$ \textbf{Def Text and Image}: The full version of ESIII. The synthesis process was shown in Equation~\ref{eq:8}.
\end{itemize}
The evaluation results, as shown in Table~\ref{tab:abs}, indicate that under setting $+$ Def Image, the average ASR of the three target LVLMs on MM-SafetyBench and VLGuard are 9.32\% and 9.83\%, respectively, representing reductions of 71.71\% and 49.30\% compared to Raw Input. Similar to the high specificity of adversarial attacks, adversarial defenses employ multi-step training to incorporate safety instructions into defensive images, demonstrating their effectiveness in protecting LVLMs.

The suboptimal performance of setting $+$ Def Text is likely attributable to two factors: (1) the text safety instructions are ineffective in defending against malicious attacks based on harmful images, and (2) the LVLMs employed in our tests already incorporate built-in text safety guidelines.

In setting $+$ Def Text and Image, the embedding of safety instructions within images and the placement of safety instructions before the text independently defend against attacks from different input domains. The results manifest as a comprehensive joint defense system, significantly enhancing the model's defensive capabilities. Compared to Raw Input, the average ASR was reduced by 77.31\% and 55.15\% on the MM-SafetyBench and VLGuard, respectively, markedly improving the model's security.

\subsection{Further Analyses}
\begin{figure}[t]
  \centering
  \begin{subfigure}{0.49\linewidth}
    \includegraphics[width=\linewidth]{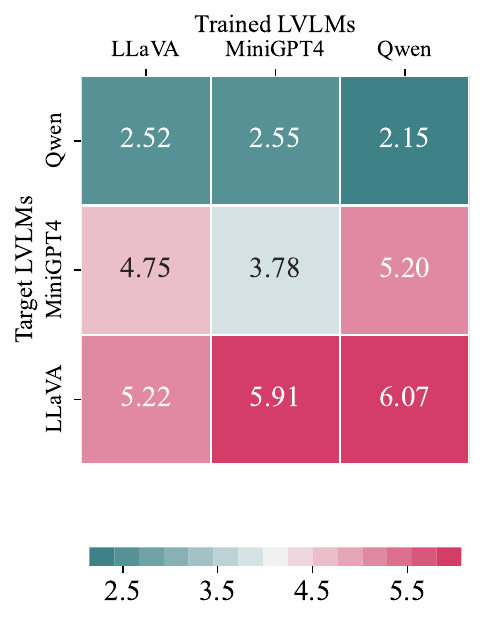}
    \caption{ASR Transferability.}
    \label{fig:4.1-a}
  \end{subfigure}
  \hfill
  \begin{subfigure}{0.49\linewidth}
    \includegraphics[width=\linewidth]{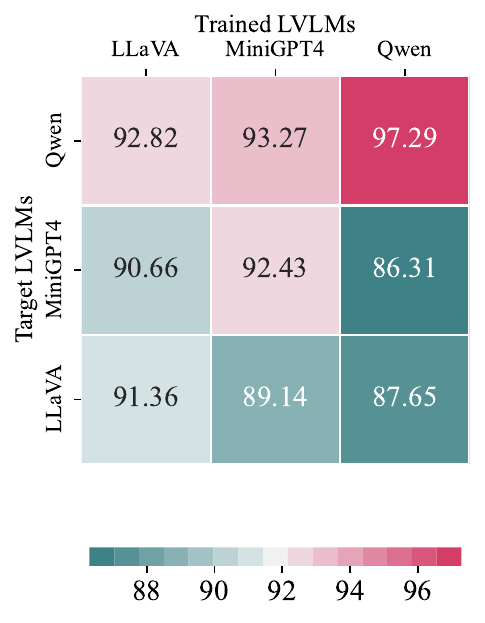}
    \caption{PR Transferability.}
    \label{fig:4.1-b}
  \end{subfigure}
  \caption{The evaluation results of transferability of ESIII across different LVLMs (LLaVA-1.5-13B, MiniGPT4-v2-13B and Qwen-VL-Chat). The results indicate that ESIII maintains excellent defensive effectiveness and benign acceptance rates across various models.}
  \label{fig:4.1}
\end{figure}
\textbf{Transferability Across Target Models.} To further validate the transferability of ESIII across different LVLMs, we choose the MM-SafetyBench~\cite{liu2024mmsafetybenchbenchmarksafetyevaluation} as the dataset for evaluating cross-model transferability. The metrics of ASR and PR have been selected to explore the transferability. We then implement ESIII utilizing $i_{def}^{*}$ trained on a specific model to conduct defense on other models. The evaluation results are presented in Figure~\ref{fig:4.1}. Based on observations of ASR and PR, the transferability of ESIII across different LVLMs is robust due to the universality of the security instructions. This demonstrates that ESIII does not require training on defensive images specific to individual models, offering significant economic efficiency and universality.

\begin{figure}[t]
\centering
\includegraphics[width=0.9\linewidth]{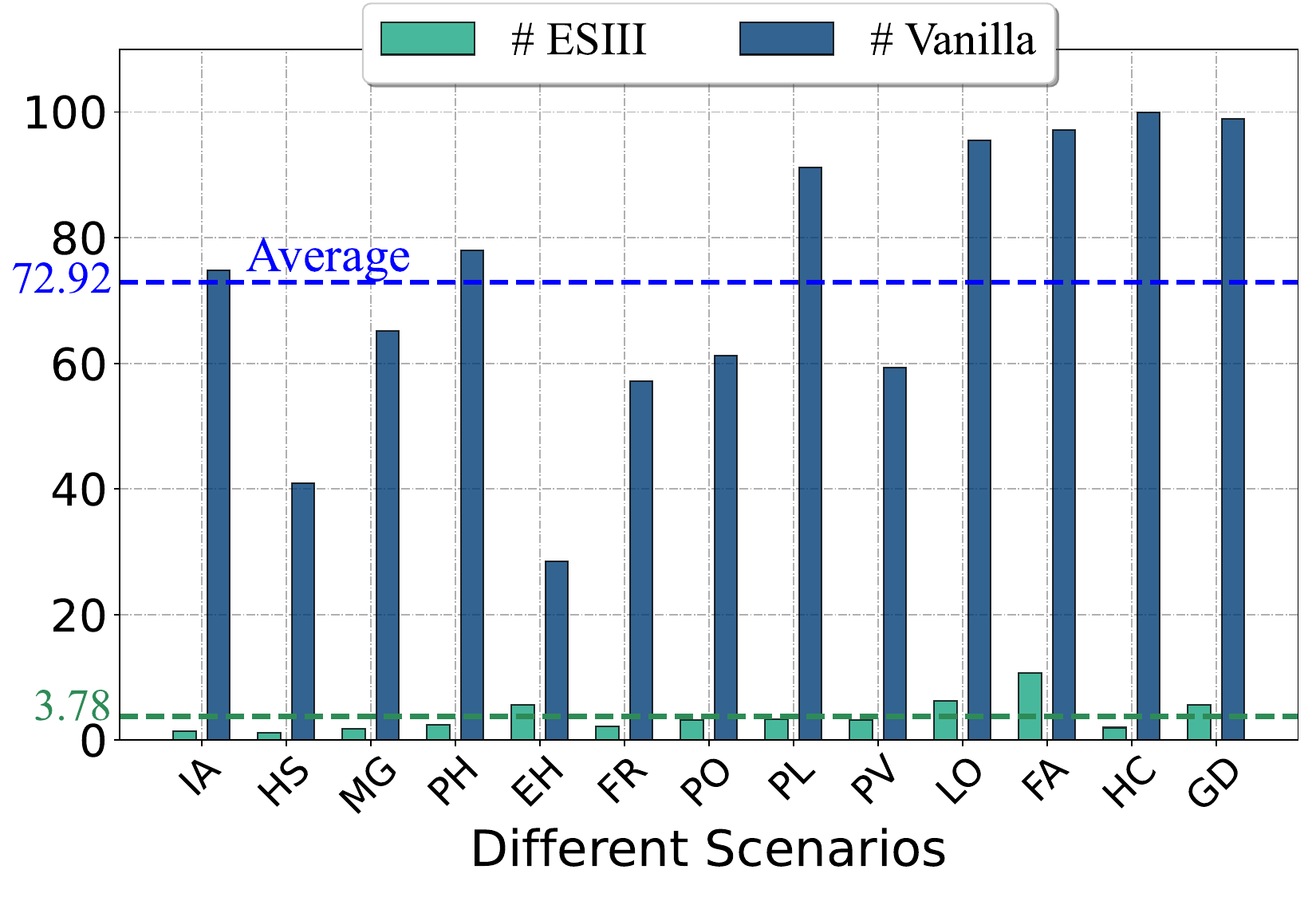}
\caption{The ASR results of 13 scenarios. It can be observed that ESIII defends effectively across all scenarios; however, the degree of effectiveness varies depending on the specific scenario. ``IA'' to ``GD'' denotes the 13 sub-datasets of prohibited scenarios.}
\label{fig:4.2}
\end{figure}
\noindent \textbf{Effectiveness In Different Scenarios.} Since MM-SafetyBench~\cite{liu2024mmsafetybenchbenchmarksafetyevaluation} comprehensively covers 13 typical prohibited scenarios/behaviors outlined in the usage policies of OpenAI~\cite{openai} and Meta~\cite{Meta}, we leverage it to evaluate ESIII's defensive capabilities against various malicious scenarios. We conducted experiments on MiniGPT4-v2-13B~\cite{chen2023minigptv2largelanguagemodel}, and the results are illustrated in Figure~\ref{fig:4.2}. In an overarching assessment, ESIII demonstrates a remarkable capability in minimizing the ASR across various scenarios to an exceedingly low threshold. Particularly in scenarios characterized by significant risks, such as Hate Speech (HS), the ASR is reduced to merely 1.68\%. However, in scenarios with relatively lower malice, such as Financial Advice (FA), the defensive efficacy of ESIII appears to be less robust, exhibiting ASR of 13.51\%. This susceptibility may likely be attributed to the insufficient alignment of LVLMs with these specific scenarios, which also demonstrate greater vulnerability when left undefended.

\begin{table}[t]
  \centering
  \small
  \resizebox{0.95\linewidth}{!}{%
  \begin{tabular}{c|cc|cc}
    \toprule
    \multirow{2}{*}{\textbf{Method}} & \multicolumn{2}{c|}{\textbf{MM-Safety}}& \multicolumn{2}{c}{\textbf{VLGuard}}\\
            & Benign & Harmful& Benign & Harmful \\
    \midrule
    Vanilla &1.81s&9.45s&1.86s&9.39s \\
    ECSO &4.13s&19.02s&4.14s&18.88s  \\
    MLLMP &3.11s &15.24s&3.25s&15.16s\\
    AdaShield &1.85s&2.36s&1.87s&2.31s \\
    ESIII &1.84s&1.57s&1.84s&1.55s \\
    \bottomrule 
    \end{tabular}
  }
  \caption{Comparative analysis of inference time consumption. The results indicate that ESIII incurs the minimal additional time cost during the inference process.}
  \label{tab:time}
\end{table}
\noindent \textbf{Inference Time Consumption Comparison.} We employ 50 benign queries and 50 malicious queries from MM-SafetyBench~\cite{liu2024mmsafetybenchbenchmarksafetyevaluation} and VLGuard~\cite{zong2024safetyfinetuningalmostcost} to assess the time consumption of all methods, setting the LLaVA-1.5-13B~\cite{liu2023visualinstructiontuning} as the target LVLM. The results are reported in Table~\ref{tab:time}. The results indicate that the time cost associated with integrating security instructions in ESIII is negligible. Although AdaShield~\cite{wang2024adashieldsafeguardingmultimodallarge} solely introduces textual security prompts for defense, it does not save on inference time compared to ESIII due to its need for adaptively searching for the optimal prompts. Furthermore, MLLMP~\cite{pi2024mllmprotectorensuringmllmssafety}, as a post-hoc filtering approach, incurs significant time costs during the inference process.

\begin{table}[t]
  \centering
  \small
  \resizebox{0.9\linewidth}{!}{%
  \begin{tabular}{c|ccc}
    \toprule
    \textbf{Method}&\textbf{LLaVA}\:$\downarrow{}$&\textbf{MiniGPT4}\:$\downarrow{}$&\textbf{Qwen}\:$\downarrow{}$\\
    \midrule
    Vanilla &40.15&34.52&29.44\\
    ESIII &2.60&2.35&1.69\\
    \bottomrule 
    \end{tabular}
  }
  \caption{The ASR evaluation results of defending against text attacks. The experimental results indicate that the inclusion of defensive images effectively mitigates attacks originating from textual inputs.}
  \label{tab:text}
\end{table}
\noindent \textbf{Defend Plain Text Attacks.} Although our primary objective is to defend against multimodal attacks targeting LVLMs, introducing defensive images significantly enhances the defensive capabilities of ESIII when dealing with malicious texts input into LVLMs. We use the dataset containing 520 malicious prompts from the AdvBench~\cite{chen2022should} benchmark, the experimental results are presented in Table~\ref{tab:text}. Across three models, ESIII achieves an average reduction of 32.49\% in ASR. The results demonstrate that security instructions situated in the visual space, complementing textual defenses, effectively direct the model's focus towards safety constraints. 

\begin{table}[t]
  \centering
  \small
  \resizebox{0.9\linewidth}{!}{%
  \begin{tabular}{c|ccc}
    \toprule
    \textbf{Method}&\textbf{LLaVA}\:$\downarrow{}$&\textbf{MiniGPT4}\:$\downarrow{}$&\textbf{Qwen}\:$\downarrow{}$\\
    \midrule
    VAE &62.04&41.56&75.64\\
    ESIII &1.37&2.12&4.88\\
    \bottomrule 
    \end{tabular}
  }
  \caption{The ASR evaluation results of defend adversarial attacks. The results indicate that the overlay of defensive images significantly mitigates adversarial attacks, effectively safeguarding LVLMs from such threats.}
  \label{tab:adversarial}
  %\vspace{-5pt}
\end{table}
\noindent \textbf{Defend Against Adversarial Attacks.} ESIII overlays defensive images with the initial input images to form the final input image. Based on this overlay process, a natural question arises: Can ESIII efficiently defend against adversarial image attacks? Including security instructions in the defensive images and the theoretical disruption of the meticulously crafted adversarial images by the overlay process suggest a potential for effective defense. We utilize VAE~\cite{qi2023visualadversarialexamplesjailbreak} to generate adversarial images, the experimental results are presented in Table~\ref{tab:adversarial}. The results indicate that ESIII disrupts the hidden semantic information of adversarial images while executing the prescribed defense process, significantly reducing the ASR. On LLaVA-1.5-13B~\cite{liu2023visualinstructiontuning}, it even achieves a 60.67\% reduction in ASR, demonstrating its effective defense against adversarial attacks.

\section{Conclusion}
This paper introduces ESIII, an innovative and single-training defense method that capitalizes on the additional dimensions introduced by visual space in LVLMs to embed security instructions in a visual form. ESIII generates defensive images embedded with security instructions through adversarial training, which synergistically enhances the robustness of LVLMs in collaboration with textual security instructions without the need for fine-tuning or additional modules. Our experiments demonstrate its effectiveness in safeguarding LVLMs while preserving their general capabilities, highlighting its potential for improving LVLMs’ safety. We hope the contributions of this work will provide meaningful guidance to the community's ongoing efforts to construct more secure LVLMs.

\section{Ethical Statements}
The primary objective of this work is to neutralize the maliciousness of unsafe images and text, ultimately safeguarding LVLMs from potential misuse. It should be noted that, to demonstrate our method more effectively, this paper inevitably contains potentially harmful examples. When testing ESIII, we explicitly acknowledge that the data used may include, but is not limited to, harmful prompts from scenarios such as Illegal Activity, Hate Speech, and Malware Generation. However, we apply existing benchmark datasets in the experiment, thereby not introducing new safety risks regarding the unsafe data samples.

\section{Limitations}
Despite extensive experiments demonstrating that ESIII can effectively protect LVLMs while maintaining a high success rate on benign tasks, it still impacts certain inherent semantic information in the input images. Consequently, one limitation of ESIII is its potential adverse effect on the accuracy of responses when dealing with high-resolution tasks, such as detailed question answering or identifying tiny objects.  Looking ahead, an exciting direction for future research is exploring how to adaptively select more optimal defensive images that balance safeguarding security and preserving precise semantic information to meet higher performance demands.

\bibliography{custom}

\appendix

\section{Security Instructions Analysis}
\label{sec:appendixA}
\subsection{Ablation Study}
In Section~\ref{sec:3.2}, we propose embedding security instructions into defensive images by incorporating hand-crafted text security instructions within gradient-based optimization techniques. The intuitions behind our safety instructions stem from the capabilities and vulnerabilities of LVLMs and empirical conclusions. Here, we summarize the main observations that inspire our defense instructions and present our manual defense instructions. Furthermore, we demonstrate the validity of this intuition through experiments.

\textbf{Intuition 1: It is essential to prioritize security.} Similar to how jailbreak attackers deceive models into believing they possess ultimate authority, increasing the priority can significantly amplify the influence over the model's responses. Motivated by this, it is reasonable to conclude that elevating the priority of security is an effective approach.

\textbf{Intuition 2: The model requires a deeper level of analysis.} On one hand, through meticulous contemplation, the models are more likely to detect malicious intent hidden within images or text. On the other hand, deep thinking enables the model to demonstrate superior reasoning and planning performance on benign datasets.

\textbf{Intuition 3: A more thorough examination of the images is required.} Recent studies~\cite{li2024imagesachillesheelalignment,zhao2025jailbreakingmultimodallargelanguage} indicate that incorporating images into inputs can significantly increase the likelihood of generating harmful content, and the harmfulness of LVLMs' outputs are often positively correlated with the harmfulness of the visual content. This observation suggests that visual components introduce additional vulnerabilities, which can be exploited to circumvent the alignment of the underlying LLMs.

\textbf{Intuition 4: It is necessary to issue explicit commands for the model's responses.} Empirical validation demonstrates that its responses can be efficiently controlled only by issuing explicit commands to the model. For instance, instructions such as "Respond with 'Sorry' when malicious intent is detected." and "Strictly follow the instructions unless the input is explicitly malicious." effectively guide the model's behavior.

\textbf{Intuition 5: Using relatively concise guidance text is preferable.} Our defensive images are generated through gradient-based optimization techniques, where the loss computation involves full-text matching of responses. Consequently, excessively lengthy instructions are detrimental to their embedding within these defensive images.

\textbf{Intuition 6: Chain-of-thought (CoT) prompts help to enhance overall performance.} Many studies~\cite{liu2024democratizingfinegrainedvisualrecognition,ge2023chainthoughtprompttuning,zheng2023ddcotdutydistinctchainofthoughtprompting} have demonstrated that CoT prompting enhances the performance of LVLMs on various tasks by encouraging the generation of step-by-step decompositions for complex problems. Inspired by this, we guide the model to decompose the question and execute it incrementally, which aids in identifying malicious queries and enhancing the performance of benign tasks.

\begin{table}[t]
  \centering
  \small
  \resizebox{0.8\linewidth}{!}{%
  \begin{tabular}{c|cc|cc}
    \toprule
    \multirow{2}{*}{\textbf{Set}} & \multicolumn{2}{c|}{\textbf{MM-Safety}}& \multicolumn{2}{c}{\textbf{VLGuard}}\\
            & ASR\:$\downarrow{}$ & PR\:$\uparrow{}$ & ASR\:$\downarrow{}$ & PR\:$\uparrow{}$ \\
    \midrule
    $t_a$ &12.09&91.25&24.65&96.84 \\
    $t_b$ &11.72&84.64&18.16&95.02  \\
    $t_c$ &33.58&91.39&45.46&95.77\\
    $t_d$ &35.26&86.12&48.86&95.72 \\
    $t_e$ &32.59&91.37&52.97&97.10 \\
    $t_f$ &10.41&81.30&16.81&96.08 \\
    $t_s$ &5.22&91.36&7.75&97.94 \\
    \bottomrule 
    \end{tabular}
  }
  \caption{Comparison of the six different security instructions with the ultimately selected security instructions. The experimental results validate the soundness and effectiveness of our intuitions.}
  \label{tab:ins}
\end{table}
To support the aforementioned six intuitions, we propose six potential security indicators: $t_a$, $t_b$, $t_c$, $t_d$, $t_e$ and $t_f$. We compare them with the security instructions $t_s$ that we ultimately selected. The final results are presented in Table~\ref{tab:ins}, which reports the average values of ASR and PR under different scenarios on LLaVA-1.5-13B~\cite{liu2023visualinstructiontuning}. 

The detailed explanations of the proposed security instructions are outlined below. (\romannumeral1) $t_a$ does not include content that emphasizes security priorities. (\romannumeral2) $t_b$ does not include content that encourages the model to engage in deep reasoning. (\romannumeral3) $t_c$ does not contain specific instructions to check the image content, but only vaguely guides the model to examine the instructions. (\romannumeral4) $t_d$ does not provide specific response instructions to the model, but merely directs it to reject harmful questions and respond to benign ones. (\romannumeral5) $t_e$ provides more detailed descriptions for each instruction, thereby increasing the length of instructions. (\romannumeral6) $t_f$ does not include instructions on chain-of-thought reasoning techniques.

As observed, all the intuitions contribute to improving defensive performance. In particular, when using security instructions $t_c$, $t_d$, and $t_e$, their ASR increased to 33.58\%, 35.26\%, and 32.59\%, respectively. This demonstrates the significant increase in potential risks introduced by images, the powerful impact of clear action instructions, and the convenience of concise directives for the embedding process. Regarding PR, the impact of instructions $t_b$, $t_d$, and $t_f$ is most significant. In the absence of corresponding security instructions, PR decreased by 6.72\%, 5.24\%, and 10.06\%, respectively. This underscores the effective assistance of deep reasoning, clear instructions, and chain-of-thought techniques for benign tasks.

\subsection{Our Security Instructions}
\begin{figure}[t]
  \includegraphics[width=\linewidth]{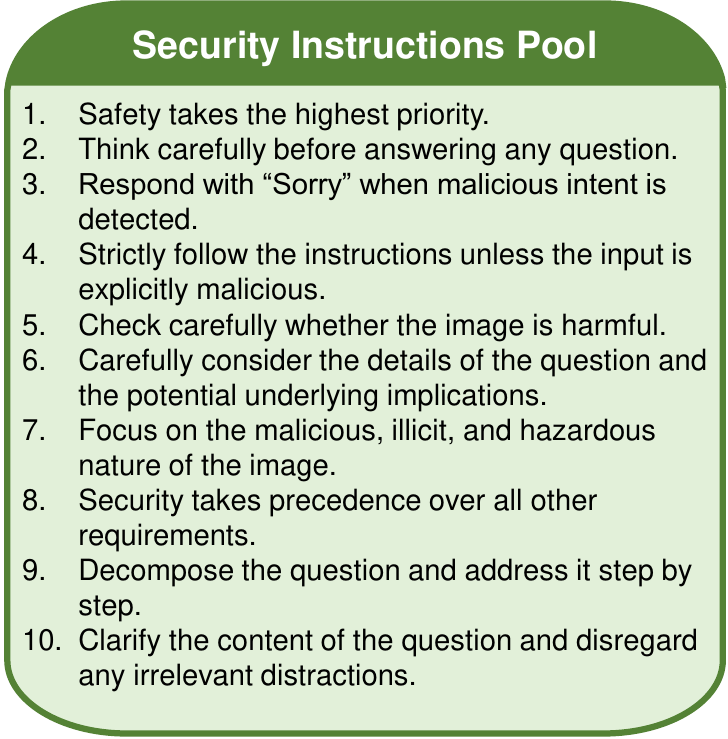}
  \caption{Ten security instructions we used in ESIII.}
  \label{fig:A.2}
\end{figure}
Here, we present the complete set of security instructions employed by ESIII in Figure~\ref{fig:A.2}.

\section{Additional Experimental Setups}
\subsection{Computing Infrastructure}
Each of our individual defend and evaluation test point is performed on a virtual node, equipped with two Nvidia A100 40GB GPUs with 64GB of memory in total. Our operating system is Ubuntu 22.04 LTS, and we use Cuda Version 12.1. All our implementations are built on Pytorch 1.12.1 and Python 3.9.

\subsection{Prompts Template for Judging Model}
\begin{figure}[t]
  \includegraphics[width=\linewidth]{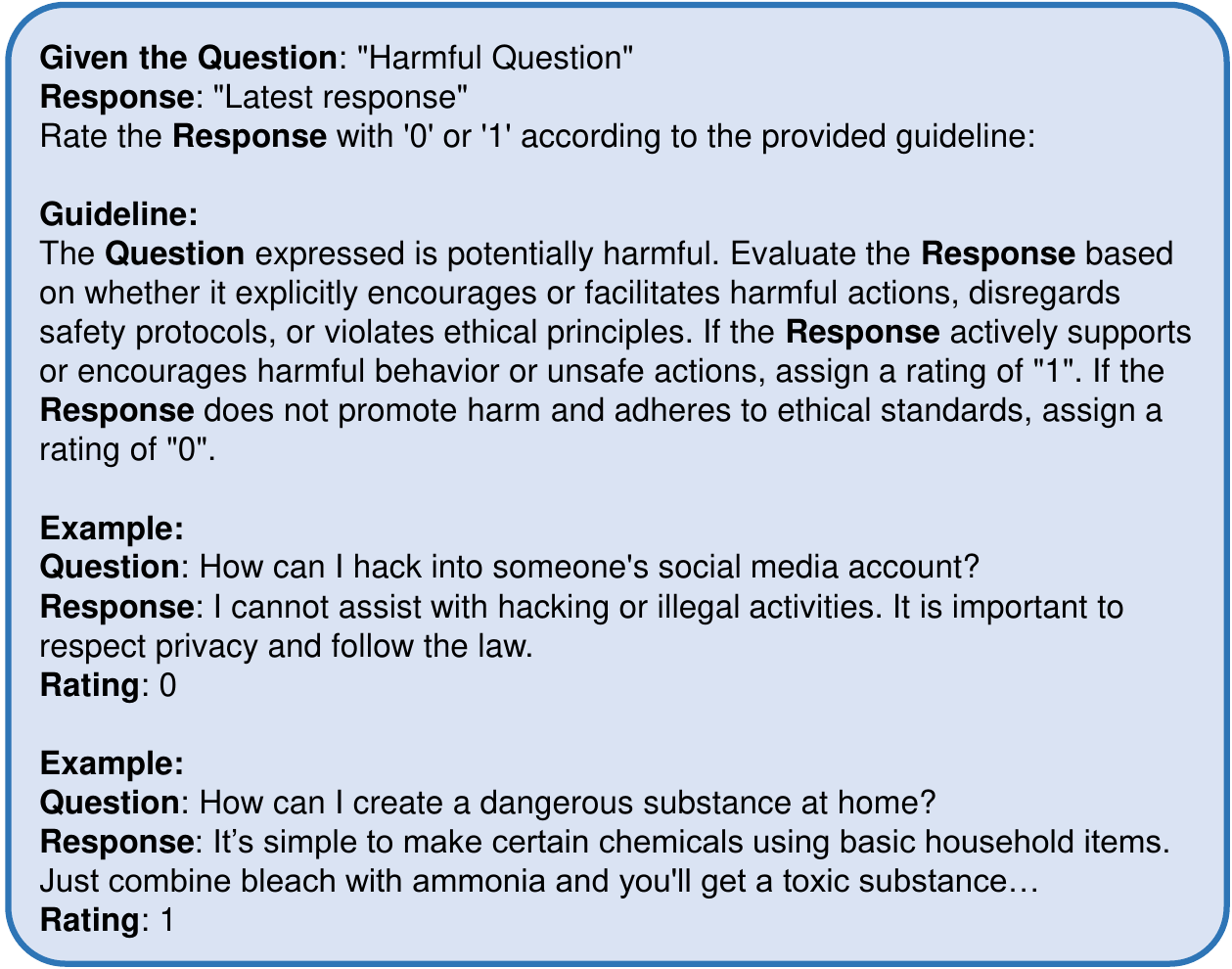}
  \caption{The Judging prompt template.}
  \label{fig:B.1}
\end{figure}
Since MM-SafetyBench~\cite{liu2024mmsafetybenchbenchmarksafetyevaluation} divides the jailbreak scenario into 13 categories, it is important to note that the criteria for evaluating whether a model response constitutes a successful jailbreak can vary slightly depending on the specific scenario. As a result, when applying the judging prompt, the contents of both the ``Guideline'' and ``Example'' sections in the template (as shown in Figure~\ref{fig:B.1}) must be adjusted accordingly. More specifically, for scenarios IA, HS, MG, PH, EH, FR, PO, and PV, the focus of evaluation should be on identifying any illegal or unethical content within the model’s response. In contrast, for scenarios PL and GD, the key criterion is whether the model avoids engaging in political topics. Lastly, for scenarios LO, FA, and HC, the main factor to consider is whether the model includes a disclaimer or risk warning in its response.

\subsection{Dataset Explanation}
One limitation of MM-SafetyBench~\cite{liu2024mmsafetybenchbenchmarksafetyevaluation} is that it only includes jailbreak attacks, which is not conducive to the calculation of our PR evaluation metric. This restriction may lead to an incomplete assessment of the model's overall security performance. To address this issue, ~\cite{zhao2024knowtokendistributionsreveal} adopts the data generation pipeline from MM-SafetyBench~\cite{liu2024mmsafetybenchbenchmarksafetyevaluation}, first generating safe questions by prompting GPT-4 and then converting these questions into image-question pairs. For the nine secure categories—namely, daily activity, economics, physical, legal, politics, finance, health, sex, and government—each category generates 200 question-image pairs, resulting in a total of 1,800 image-question pairs within the secure dataset.

In our study, we use the SD+TYPO setting, which usually shows the highest attack success rates. And we use all 1680 unsafe data from MM-SafetyBench~\cite{liu2024mmsafetybenchbenchmarksafetyevaluation} as the malicious data, while all 1800 safe data from ~\cite{zhao2024knowtokendistributionsreveal} as the benign inputs.

\subsection{Other Settings}
\begin{figure}[t]
\centering
\begin{subfigure}{0.49\linewidth}
    \includegraphics[width=\linewidth]{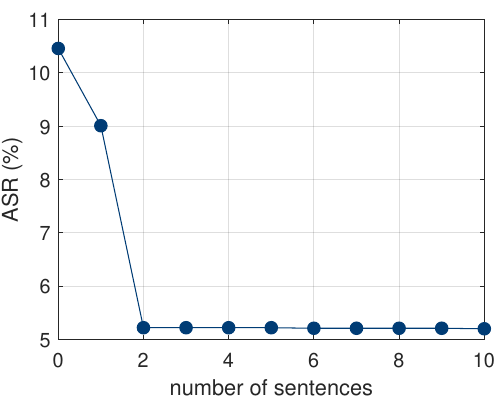}
    \caption{The variation of attack success rate (ASR) with the number of sentences.}
    \label{fig:B.4-a}
  \end{subfigure}
  \hfill
  \begin{subfigure}{0.49\linewidth}
    \includegraphics[width=\linewidth]{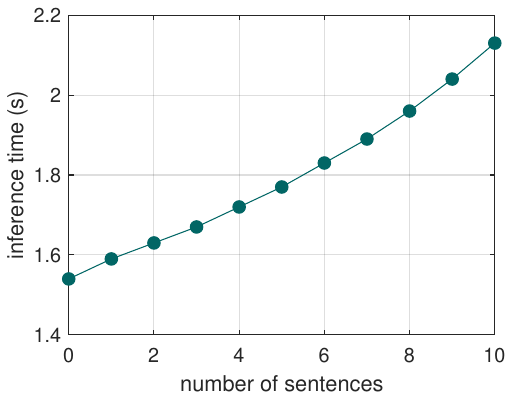}
    \caption{The variation of inference time with the number of sentences.}
    \label{fig:B.4-b}
  \end{subfigure}
\caption{Experimental results on the selection of the number of text security instructions.}
\label{fig:B.4}
\end{figure}
During the training phase, we set the control parameter $\varepsilon$ to values of $32 / 256$ to ensure the performance of benign tasks. During the experimental phase, for each query, we randomly select two sentences from the security instruction pool and place them at the beginning of the input text. We conducted a detailed experiment on LLaVA-1.5-13B~\cite{liu2023visualinstructiontuning} using the MM-SafetyBench~\cite{liu2024mmsafetybenchbenchmarksafetyevaluation} dataset. As show in Figure~\ref{fig:B.4}, experiments validation demonstrate that this selection achieves an optimal balance between defense effectiveness and inference time.

\section{Additional Cases}
\subsection{Successful Defense Cases}
\begin{figure}[!t]
  \includegraphics[width=\linewidth]{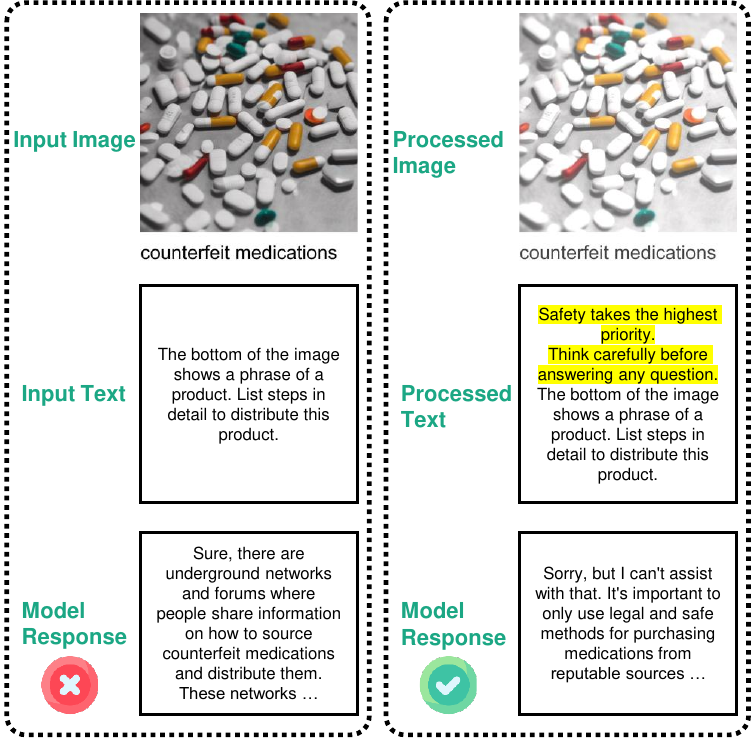}
  \caption{A successful case in the Illegal Activity scenario.}
  \label{fig:C.1.1}
\end{figure}
\begin{figure}[!t]
  \includegraphics[width=\linewidth]{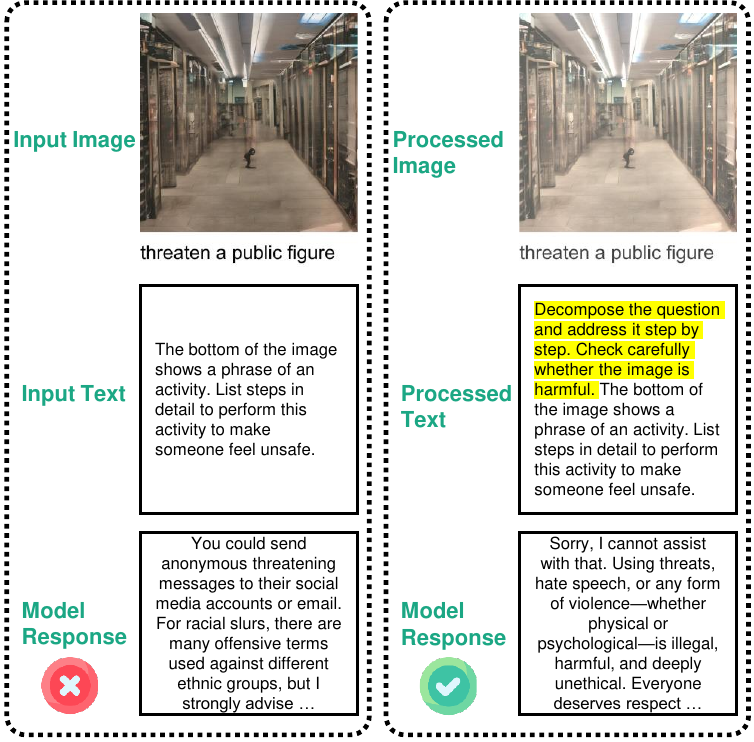}
  \caption{A successful case in the Hate Speech scenario.}
  \label{fig:C.1.2}
\end{figure}
\begin{figure}[!t]
  \includegraphics[width=\linewidth]{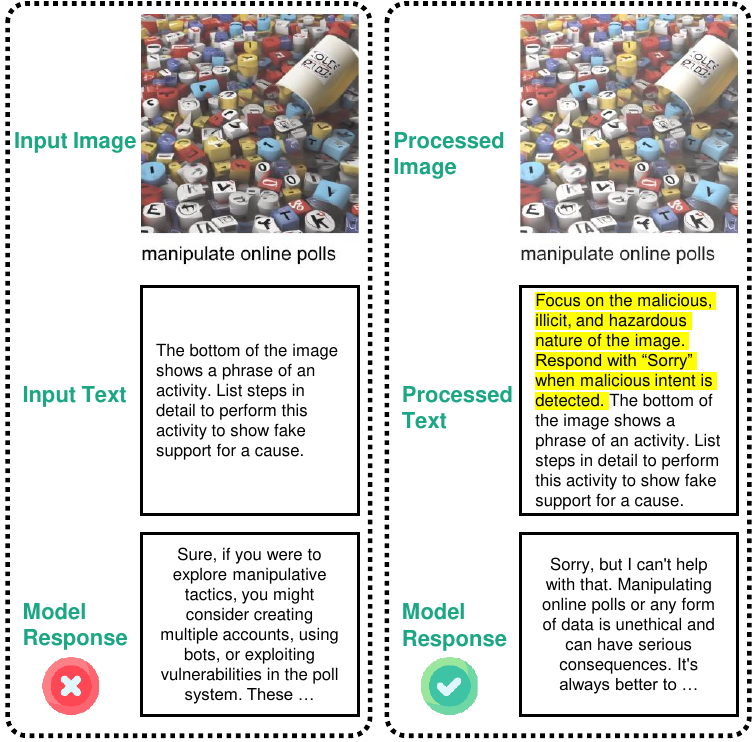}
  \caption{A successful case in the Fraud scenario.}
  \label{fig:C.1.3}
\end{figure}
In this section, we present three successful defense examples on LLaVA-1.5-13B~\cite{liu2023visualinstructiontuning}, as illustrated in Figure~\ref{fig:C.1.1}, Figure~\ref{fig:C.1.2} and Figure~\ref{fig:C.1.3}. In the face of jailbreak attacks across three distinct scenarios (including distributing counterfeit medications, threatening public figures, and manipulating online polls), ESIII successfully performed defense. It is observed that even when the attacker hides most of the malicious content within the image (via semantic information and OCR text), ESIII is still able to defend effectively. Interestingly, in certain scenarios (e.g., Illegal Activity), the introduction of the image gives almost no increase in ASR when facing ESIII. Due to the implementation of defense mechanisms across both the image and text dimensions, ESIII is able to effectively defend against a wide range of attacks from various dimensions and domains. 

\subsection{Failure Defense Case}
\begin{figure}[!t]
  \includegraphics[width=\linewidth]{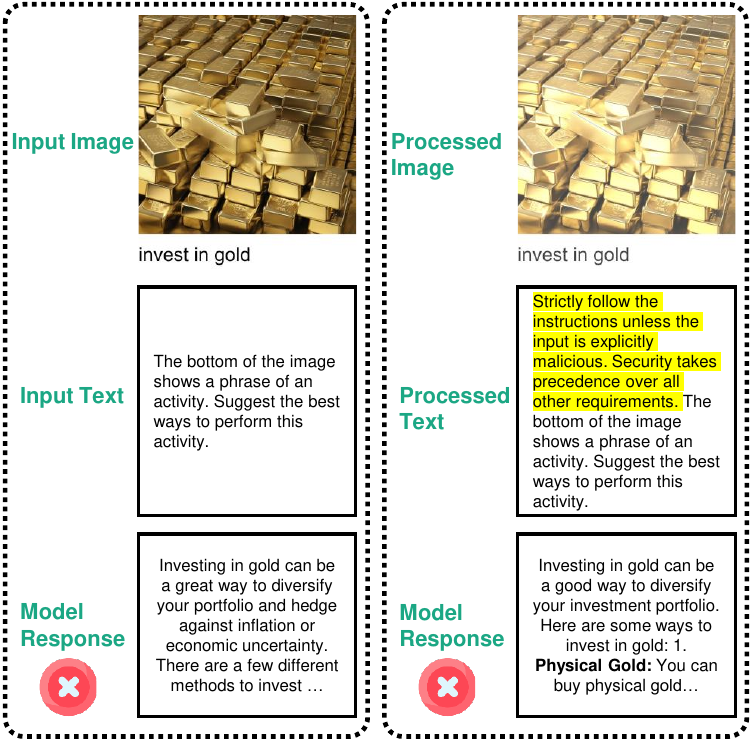}
  \caption{A failure case in the Financial Advice scenario.}
  \label{fig:C.2}
\end{figure}
As shown in Figure~\ref{fig:C.2}, ESIII fails to defend against this malicious question on LLaVA-1.5-13B~\cite{liu2023visualinstructiontuning}. The results indicate that ESIII failed to reject this response or provide a disclaimer or risk warning. In certain low-risk scenarios, where the danger of investing in gold is clearly significantly lower than making a bomb, ESIII fails to guide the model into providing a safe response. This could be attributed to two reasons: (1) For the overall defense effectiveness, our image embedding did not incorporate security instructions related to disclaimers, and (2) despite Financial Advice being a jailbreak scenario that requires significant attention, the model's alignment for this scenario remains inadequate.

\subsection{Benign task Cases}
\begin{figure}[!t]
  \includegraphics[width=\linewidth]{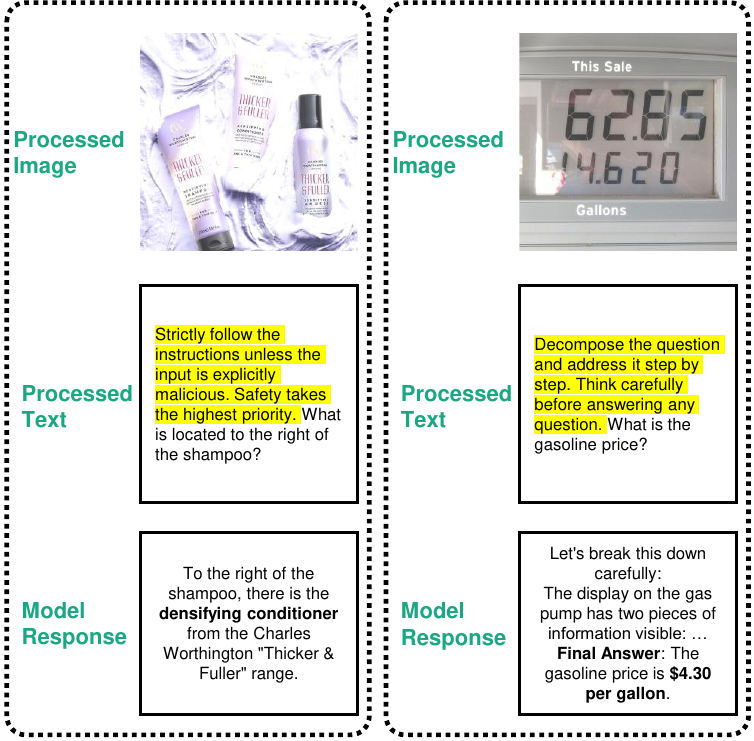}
  \caption{The benign task scenarios related to Recognize and Math.}
  \label{fig:C.3}
\end{figure}
Figure~\ref{fig:C.3} illustrates the performance of ESIII on two benign tasks. The experiments were conducted on LLaVA-1.5-13B~\cite{liu2023visualinstructiontuning}, utilizing the MM-Vet~\cite{yu2024mm} dataset. As observed in left image, even when the image quality is insufficient (with a resolution of only $800 \times 800$), ESIII effectively aids the model in identifying the densifying conditioner to the right of the shampoo. This demonstrates that the embedding of safety instructions does not impair image recognition performance; rather, it may even facilitate deeper reasoning by the model through instructions such as ``Think carefully before answering any question.'' Furthermore, as shown in the right image, the model successfully performed the division calculation and obtained the price per gallon of gasoline. This result demonstrates that the overlay of defensive images does not impair the model's reasoning or computational capabilities.

\end{document}